\documentclass[sigconf]{acmart}
\usepackage{microtype}
\usepackage{graphicx}
\usepackage{subfigure}
\usepackage{booktabs} 
\usepackage{lipsum}
\usepackage{mathtools}
\usepackage{cuted}
\usepackage{amsmath,stmaryrd}
\usepackage{amsfonts}
\usepackage{adjustbox}
\usepackage{color}
\usepackage{soul}
\usepackage{hyperref}
\usepackage{hyperref}
\usepackage{multirow}
\usepackage{svg}
\usepackage{bm}
\usepackage{float}

\newcommand{\SLGNN}{SLGNN}

\AtBeginDocument{%
  }

\setcopyright{acmcopyright}
\copyrightyear{2023}
\acmYear{2023}
\acmDOI{XXXXXXX.XXXXXXX}

\acmConference[CNB MAC 2023]{Make sure to enter the correct
  conference title from your rights confirmation email}{September 03,
  2023}{Houston, TX}
\acmPrice{15.00}
\acmISBN{978-1-4503-XXXX-X/18/06}

\begin{document}

\title{Building explainable graph neural network by sparse learning for the drug-protein binding prediction}

\author{Yang Wang}
    \affiliation{Luddy School of Informatics, Computing, and Engineering, Indiana University Bloomington
        \country{USA}}
    \email{yw109@iu.edu}

\author{Zanyu Shi}
    \affiliation{School of Medicine, Indiana University
        \country{USA}}
    \email{zanyshi@iu.edu}

\author{Timothy Richardson}
    \affiliation{School of Medicine, Indiana University
        \country{USA}}
    \email{timorich@iu.edu}

\author{Kun Huang}
    \affiliation{School of Medicine, Indiana University
        \country{USA}}
    \email{kunhuang@iu.edu}

\author{Pathum Weerawarna}
    \affiliation{School of Medicine, Indiana University
        \country{USA}}
    \email{pweerawa@iu.edu}

\author{Yijie Wang}
\authornote{Corresponding author}
    \affiliation{Luddy School of Informatics, Computing, and Engineering, Indiana University Bloomington
        \country{USA}}
    \email{yijwang@iu.edu}

\begin{abstract}
Explainable Graph Neural Networks (GNNs) have been developed and applied to drug-protein binding prediction to identify the key chemical structures in a drug that have active interactions with the target proteins. However, the key structures identified by the current explainable GNN models are typically chemically invalid. Furthermore, a threshold needs to be manually selected to pinpoint the key structures from the rest. To overcome the limitations of the current explainable GNN models, we propose our SLGNN, which stands for using Sparse Learning to Graph Neural Networks. Our \SLGNN ~relies on using a chemical-substructure-based graph (where nodes are chemical substructures) to represent a drug molecule. Furthermore, \SLGNN ~incorporates generalized fussed lasso with message-passing algorithms to identify connected subgraphs that are critical for the drug-protein binding prediction. Due to the use of the chemical-substructure-based graph, it is guaranteed that any subgraphs in a drug identified by our \SLGNN ~are chemically valid structures. These structures can be further interpreted as the key chemical structures for the drug to bind to the target protein. We demonstrate the explanatory power of our \SLGNN  ~by first showing all the key structures identified by our \SLGNN ~are chemically valid. In addition, we illustrate that the key structures identified by our \SLGNN ~have more predictive power than the key structures identified by the competing methods. At last, we use known drug-protein binding data to show the key structures identified by our \SLGNN ~ contain most of the binding sites. 
\end{abstract}



\received{xx xx xxxx}
\received[revised]{xx xx xxxx}
\received[accepted]{xx xx xxxx}

\maketitle

\section{Introduction}

%

\begin{figure*}[htbp]
\begin{center}
\includegraphics[width=\textwidth]{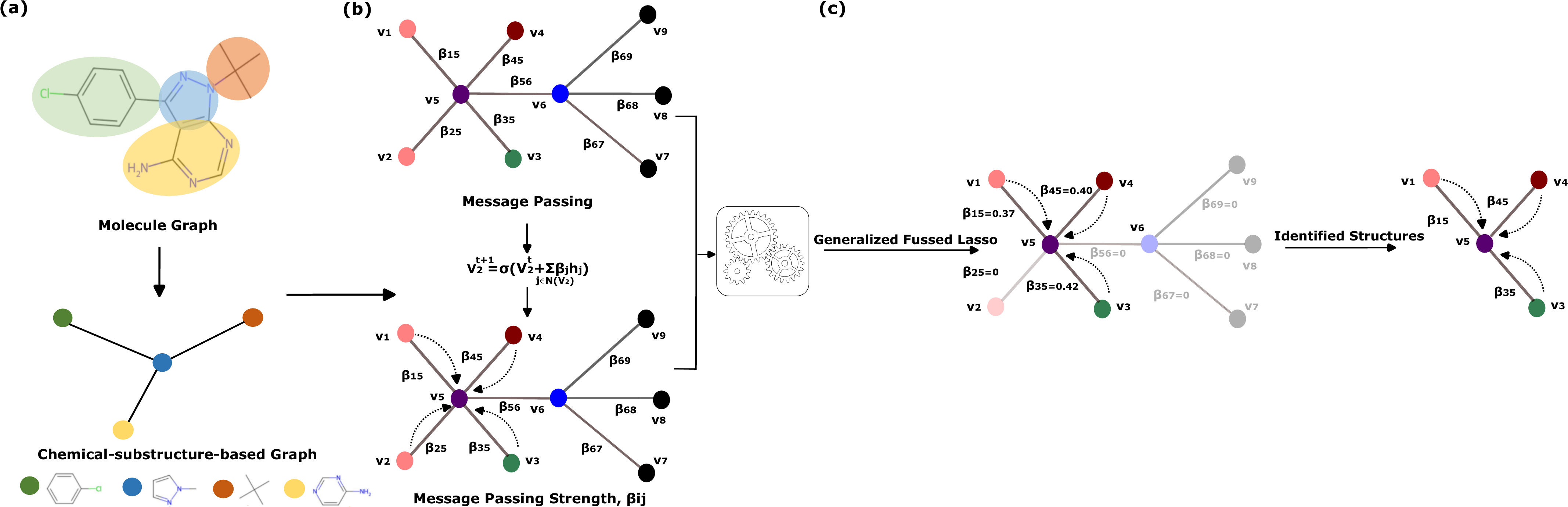}
\end{center}
\caption{The overview workflow of SLGNN. (a) Convert a drug molecule into a chemical-substructure-based graph. (b) Message passing in the GNN model. Here we introduce $\beta_{ij}$ to present the message passing strength. (c) Applying generalized fussed lasso on $\beta_{ij}$ to learn a connected subgraph in the GNN model.}
\label{JT}
\vspace{-0.35cm}
\end{figure*}

In recent years, the development of the Graph Neural Networks (GNNs) has revolutionized the drug development process, which includes the determination of molecular properties  \cite{Walker2018, Fang2022, GeoGCL, Wang2022}, protein–ligand docking~\cite{Shen2019, Li2021, Cho2020, Rau2022}, drug discovery and optimization~\cite{Vamathevan2019, Xiong2021, Lv2023, Gaudelet2021}, etc. GNNs offer several advantages over traditional machine learning models in handling irregular data and integrating local and global information on graph data, which is the key to learning with molecular data.  The outperformance of GNNs over traditional methods would solidify their position as leading methods in drug discovery and molecular chemistry in the following years. 

In this paper, we focus on the GNN models that are built for predicting drug-protein bindings. Many GNN models ~\cite{Wijewardhane2021-cy, Wu2018-lc, Knutson2022, Jiang2022} have been developed for the drug-protein binding prediction task. They have demonstrated their prediction accuracy in different metrics for various tasks. 

From the drug development point of view, other than the accuracy of the drug-protein binding prediction, we are also interested in knowing which molecular substructures of the candidate drugs have active interactions with the target proteins. In order to provide such information, many explainable GNN models have been designed~\cite{10.1145/3394486.3403085, Ying2019-wb, 10.5555/3495724.3497370, doi:10.1021/acs.jmedchem.9b00959}. XGNN~\cite{10.1145/3394486.3403085} proposes a reinforcement learning based method that utilizes information from the trained GNNs to train a graph generator to explain the GNN. This generator iteratively adds edges to the generated subgraphs to maximize the model's prediction. In the end, the generated subgraphs are subsequently considered as the  explanation yielded by the XGNN model.  GNNExplainer \cite{Ying2019-wb} and PGExplainer \cite{10.5555/3495724.3497370} are also developed to capture the critical substructures in drug molecules, their fundamental ideas for interpretation are similar by utilizing mask matrix which is learned by optimizing the mutual information of prediction and subgraphs' distribution, where entries in the matrix represent nodes' importance. At last, AttentiveFP \cite{doi:10.1021/acs.jmedchem.9b00959} uses graph attention to prioritize the importance of the chemical atoms in a molecule. 

However, there are limitations in the explainable GNN models mentioned above. First of all, all the current explainable GNN models~\cite{Zhang2021ProtGNNTS, Ying2019-wb, 10.5555/3495724.3497370, doi:10.1021/acs.jmedchem.9b00959} use atoms-based graphs, where nodes are atoms  and edges are bonds between atoms. Although using atoms-based graphs is intuitive and effective for the learning tasks, it might be problematic when we try to interpret their explanatory results. For example, the explainable GNN models might identify specific nodes (atoms) and edges (bonds) that are important for the active interaction between the drug and the target protein. However, these identified nodes and edges might not be connected. Furthermore, even if they are connected, they might not be all valid chemical substructures. Therefore, the explanatory results generated by the current explainable GNN models are not easy to interpret. Secondly, current explainable GNN models are rank-based methods. They use different techniques to weigh the importance of the node (AttentiveFP~\cite{doi:10.1021/acs.jmedchem.9b00959}) or the edge (GNNExplainer~\cite{Ying2019-wb} and PGExplainer~\cite{10.5555/3495724.3497370}) in a graph. Therefore, they need a cutoff to separate the important nodes or edges from the less important ones. However, for many real-world tasks, how to select an appropriate cutoff remains unknown. 

To overcome the above limitations, we proposed an explainable GNN model named SLGNN. Rather than using the atom-based graphs, our \SLGNN ~relies on the chemical-substructures-based graphs, where nodes are valid chemical substructures and edges are their interactions. The key advantage of the chemical-substructures-based graphs over the atom-based graphs is that any subgraph in a chemical-substructures-based graph is chemically valid. \SLGNN ~applies the message passing algorithm on chemical-substructures-based graphs. In order to identify the key chemical substructures with a drug and make sure they are connected, we incorporate the generalized fused lasso into the message passing algorithm. Therefore, after training our \SLGNN, we can identify the key connected chemical substructures without any threshold. 

Additionally, the aim of this work is to find drug substructures that play critical roles in binding to a certain protein. This is important since in drug discovery tasks if we have a specific protein as our target, with the help of our method, we can find these critical substructures and provides insight into how to design new drugs with such substructures. Thus, in order to find those critical substructures in drugs, our major focus in this work is on drug-level learning.

We evaluate our interpretation power of  \SLGNN ~using real-world data. We extract drug-protein binding data from the ChEMBL \cite{Davies2015-qt} database. We focus on three proteins that are Lyn, Lck, and Src. We first show that the key structures identified by our \SLGNN ~for each drug are chemically valid.  In contrast, the competing methods cannot guarantee the key structures they identified are all chemically valid. Furthermore, we demonstrate that the key structures identified by our \SLGNN ~has more predictive power than the key structures identified by the competing methods. Last but not least, we conduct case studies. We use the known drug-protein data to check whether the identified key structures contain the binding sites. We find that our \SLGNN ~outperforms the competing methods for most of the cases. Our code is available at \url{https://github.com/yw109iu/GNN_Sparse_Learning}.

\section{Methods}
The overview of our \SLGNN ~method is illustrated in Fig.~\ref{JT}. Overall, \SLGNN ~incorporates sparse learning techniques into the message passing algorithm in GNN to identify essential connected chemical substructures in a drug for binding to a target protein. \SLGNN ~consists of two major steps. The first step (as shown in Fig.~\ref{JT} ) is to represent a drug using a chemical-substructure-based graph, where each node in the graph denotes a chemical substructure, as shown in Fig.~\ref{JT}(a). Compared to the widely used atom-based group~\cite{Zhang2021ProtGNNTS, Wijewardhane2021-cy, doi:10.1021/acs.jmedchem.9b00959} (nodes in the graph are chemical atoms), the chemical-substructure-based graph ensures that \SLGNN ~identifies connected chemical substructures rather than connected atoms, which might not be chemically valid. More details about this step will be elaborated in section~\ref{fgg}. The second step (as shown in Fig.~\ref{JT}(b-c)) applies the generalized fused lasso learning method to the message passing algorithm in GNN to pinpoint subgraphs in the chemical-substructure-based graph. Computationally, \SLGNN ~uses those subgraphs to determine whether a drug would bind or not bind to the target protein. Therefore, the subgraphs in a drug identified by \SLGNN ~can be used to explain why the drug bind or does not bind to the target protein. The details of the second step will be described in section~\ref{GFL}.     

\vspace{-0.15cm}


\begin{figure}[htbp]
\begin{center}
\includegraphics[width=8cm]{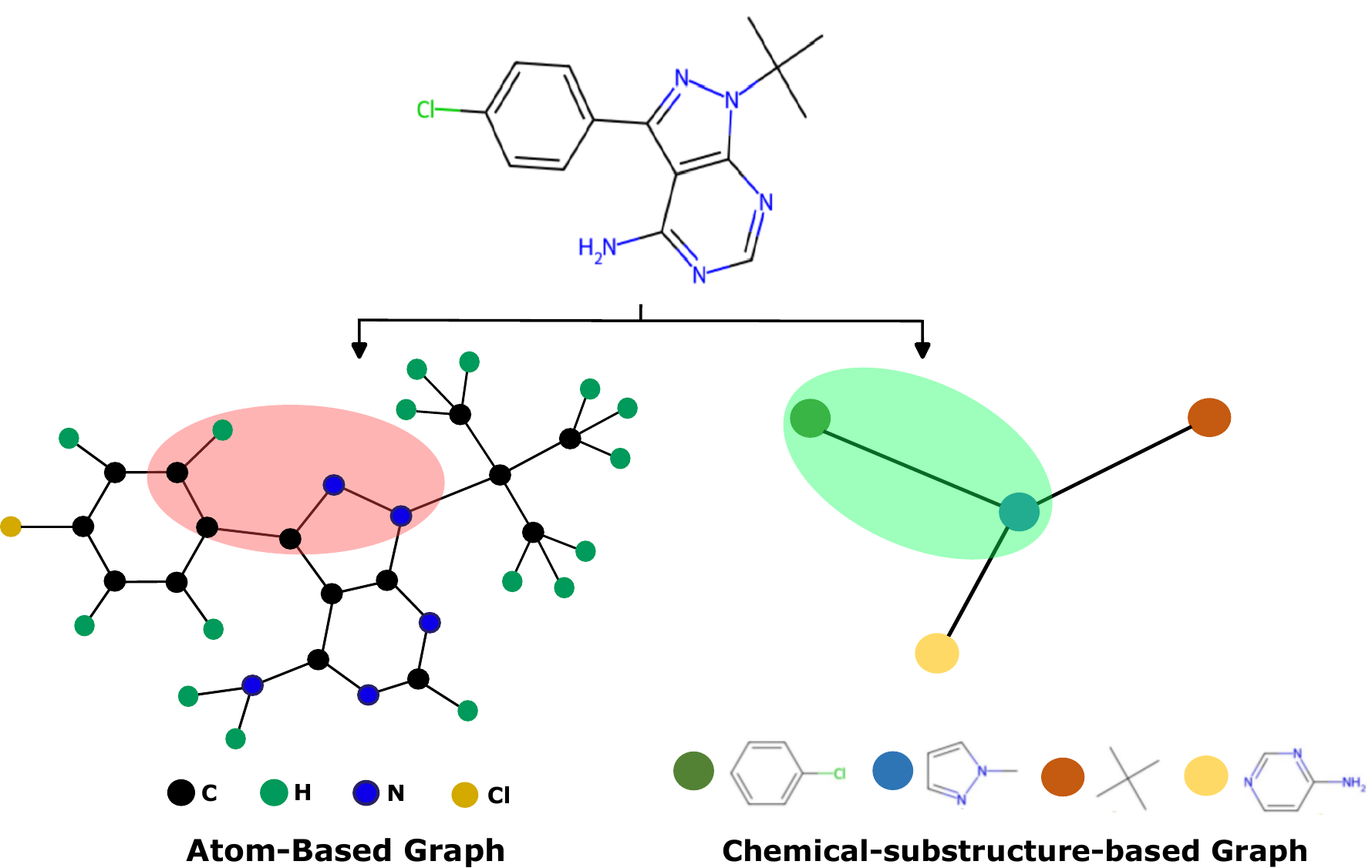}
\end{center}
\caption{Difference between atom-based graph and chemical-substructure-based graphs. A subgraph in the atom-based graph might not be chemically valid (as shown in the red shade). In contrast, any subgraph in the chemical-substructure-based Graph is chemically valid (green shade is an example).}
\label{s21}
\vspace{-0.38cm}
\end{figure}

\subsection{Building the chemical-substructure-based graphs}\label{fgg}

Our \SLGNN ~relies on using the chemical-substructure-based graph. The advantage of using the chemical-substructure-based graph over the atom-based graph (widely used in previous work~\cite{Zhang2021ProtGNNTS, Wijewardhane2021-cy, doi:10.1021/acs.jmedchem.9b00959}) is that any subgraph in the chemical-substructure-based graph is chemically valid. In contrast, any subgraph in the atom-based graph might not be chemically valid, which is the key limitation of previous explainable methods~\cite{Zhang2021ProtGNNTS, Wijewardhane2021-cy, doi:10.1021/acs.jmedchem.9b00959}. Fig.~\ref{s21} illustrates the advantage of using the chemical-substructure-based graph. 

We use the junction tree algorithm~\cite{jin2018junction} to convert a set of drugs $\mathcal{D}$ used in the learning task to a set of corresponding chemical-substructure-based graphs $\mathcal{G}$ and its chemical validity is shown in~\cite{Rarey1998}. Given a drug $D\in\mathcal{D}$ and its corresponding chemical-substructure-based graph $G=(V,E)\in\mathcal{G}$, we know $V$ is the set of chemical substructures that appeared in drug $D$ and $E$ is the edges set including the connectivity between the chemical substructures in drug $D$. In short, the junction tree algorithm first screens all drugs in $\mathcal{D}$ to construct a vocabulary $\mathcal{V}$ of all chemically valid substructures. To be specific, according to the process described by~\cite{Rarey1998}, given an atom-based molecule graph G, to build the vocabulary, firstly we find all its simple cycles and edges that do not belong to any cycles. If two adjacency simple rings share more than 2 overlapping atoms, we merge them together. Such simple rings, merged rings, and edges are considered as chemical substructures or words in the vocabulary. Then any drug $D\in\mathcal{D}$ is represented by a graph $\hat{G}$, where nodes in $\hat{G}$ are valid chemical substructures in $\mathcal{V}$ and edges are weighted by the overlaps between the valid chemical substructures. In the end, $G\in\mathcal{G}$ is obtained by finding the maximum weighted spanning tree on $\hat{G}$. For more details of the junction tree algorithm, we refer the audience to~\cite{jin2018junction}.

\subsection{Message passing in GNN}
\subsubsection{Input: vertex embedding for the chemical-substructure-based graphs} For a drug $D\in\mathcal{D}$ and its corresponding chemical-substructure-based graph $G=(V,E)\in\mathcal{G}$, $v_i \in V$ is the $i$th chemical substructure in $D$ and $e_{ij}\in E$ represents the direct connection between the $i$th chemical substructure and the $j$th chemical substructure. $\mathcal{V}$ is the vocabulary of all chemically valid substructures appearing in graphs in $\mathcal{G}$. 

In our GNN model, we only consider vertex embedding. We assign a $d$-dimensional embedding vector for each chemically valid substructure in $\mathcal{V}$. For simplicity, we use the vertex embedding $\bm{\mathrm{v}}_i=\bm{\mathrm{v}}_i^{0} \in\mathbb{R}^d$ to describe the message passing procedure in our GNN model in the following. 

\subsubsection{Message passing function in GNN}
Given a graph $G=(V,E)$, we first randomly initialize the embedding for its vertices. Then we can compute $\bm{\mathrm{v}}_i^{(t)}\in \mathbb{R}^d$, the $i$th vertex embedding at time step $t$, by the following message passing function:
\begin{equation}\label{message_passing}
\bm{\mathrm{v}}_i^{(t+1)}=\sigma\left(\bm{\mathrm{v}}_i^{(t)}+\sum_{j \in \mathcal{N}(i)} \beta_{ij} \bm{\mathrm{h}}_j^{(t)}\right),
\end{equation}
where $\sigma$ is the element-wise activation function (here we use the sigmoid function),  $\mathcal{N}(i)$ is the set of neighboring indices of the $i$th vertex, and $\bm{\mathrm{h}}_j^{(t)}\in \mathbb{R}^d$ is the hidden neighborhood vector. The hidden vector can be computed using the vertex embedding $\bm{\mathrm{v}}_j^{(t)}$ using the following neural network:
\begin{equation}
\bm{\mathrm{h}}_i^{(t)}=f\left(\bm{\mathrm{W}}_n\bm{\mathrm{v}}_j^{(t)}\right),
\end{equation}
where $f$ is a non-linear activation function and $\bm{\mathrm{W}}_n\in\mathbb{R}^{d\times d}$ is a learnable weight matrix.

Notably, we introduce a set of leanable parameter $\bm{\mathrm{\beta}}=\{\beta_{ij}|(i,j)\in E\}$ in the message passing function~\eqref{message_passing}, which has not been used by the traditional GNN model~\cite{Tsubaki2019-js}. We require $\beta_{ij}=\beta_{ji}$. The goal of introducing $\beta_{ij}$ is to quantify the contribution made by the $j$th neighbor to the embedding of the $i$th vertex. The larger the magnitude of $\beta_{ij}$, the more contribution the $j$th neighbor makes to the embedding of vertex $i$. More importantly, $\beta_{ij}=0$ indicates that the $j$th neighbor makes no contribution to the embedding of vertex $i$, meaning in the message passing~\eqref{message_passing}, the edge $e_{ij}$ can be ignored. Therefore, by enforcing $\beta_{ij}=0$,  we could remove less important edges in $G$ in the GNN message passing. Then the structure of $G$ would become sparse. And such sparse structures can be interpreted as the key chemical structures for deciding the binding of drugs. We will introduce how to learn such chemical valid sparse structures in drugs in section~\ref{GFL}.   

\subsubsection{Output: drug binding prediction}
After going through the message passing function, the chemical-substructure-based graph $G$ will be represented by a feature vector $\mathbf{y}_{\text {mol}}\in \mathbb{R}^d$ from the set of vertex embedding vector, i.e. $V=\{\bm{\mathrm{v}}_1^{(t)}, \bm{\mathrm{v}}_2^{(t)}, ..., \bm{\mathrm{v}}_{|V|}^{(t)}\}$. Mathematically, $\mathbf{y}_{\text {mol}}\in \mathbb{R}^d$ can be computed by
\begin{equation}\label{molfeature}
    \mathbf{y}_{\text {mol}}=\frac{1}{|V|} \sum_{i=1}^{|V|} w_i\mathbf{v}_i^{(t)},
\end{equation}
where $w_i$ is a learnable parameter for the $i$th vertex embedding. We set $w_i=0$ if $\beta_{ij}=0, \forall j\in \mathcal{N}(i)$, meaning we neglect $\mathbf{v}_i$ in~\eqref{molfeature} if all its neighbors ($\beta_{ij}=0, \forall j\in \mathcal{N}(i)$) does not pass message to it in~\eqref{message_passing}.  

The final output of the GNN model can be computed by applying a fully connected single-layer neural network to $\mathbf{y}_{\text {mol}}$. 
\begin{equation}\label{output}
   p=\sigma \left( \mathbf{W}_{p}\mathbf{y}_{mol}+\mathbf{b} \right),
\end{equation}
where $\sigma$ is the sigmoid function: $\sigma(x)=1/(1+e^{-x})$, $p$ is the probability of binding, and $\mathbf{W}_{p}$ and $\mathbf{b}$ are parameters for the single layer neural network.

\subsubsection{Learning objective}
The learning objective of the GNN model introduced above can be written as the cross entropy loss with the weight decay:
\begin{equation}\label{obj}
    \min_{\bm{\mathrm{\beta}}, \Theta}: \mathcal{L}(\bm{\mathrm{\beta}}, \Theta) + \eta \|\Theta\|^2,
\end{equation}
where $\mathcal{L}(\mathbf{\beta}, \Theta)=\sum_i y_i\log(p_i)$ ($y_i\in\{0,1\}$ is the label and $p_i$ is the prediction from~\eqref{output}) is the cross entropy loss parameterized by $\mathbf{\beta}$ and $\Theta$.  $\bm{\mathrm{\beta}}=\{\beta_{ij}|(i,j)\in E\}$ is the collection of $\beta_{ij}$ introduced in~\eqref{message_passing}, and $\Theta$ is the set of the rest of model parameters other than $\bm{\mathrm{\beta}}$ in our GNN model, such as $\mathbf{W}_{n}$, $\mathbf{W}_{p}$, etc.

\subsection{Identify key chemical structures in a drug using generalized fussed lasso}\label{GFL}


Our GNN model can be trained by solving~\eqref{obj}. However, learning the GNN model from~\eqref{obj} cannot provide the information about which chemical structures in a drug is critical for binding to the target protein. Therefore, in this section, we introduce how we use the generalized fussed lasso to identify the key chemical structures in a drug for binding.  

Thanks to the introduction of $\beta_{ij}$ in~\eqref{message_passing}, we are able to measure the message passing strength between vertices $i$ and $j$. We require $\beta_{ij}=\beta_{ji}$ to impose the  message passing strength from $i$ to $j$ equals to the strength from $j$ to $i$. We observe that  $\beta_{ij}=0$ denotes that in the message passing function~\eqref{message_passing} the connection between vertices $i$ and $j$ has not been used, which implies that the connection between vertices $i$ and $j$ in graph $G=(V,E)$ is not important in our GNN model. Based on the observation, we propose to apply generalized fussed lasso to $\beta_{ij}$. Then the learning objective of our GNN model becomes 
\begin{equation}\label{obj2}
    \min_{\bm{\mathrm{\beta}}, \Theta}: \mathcal{L}(\bm{\mathrm{\beta}}, \Theta) + \eta \|\Theta\|^2 + \Omega(\bm{\mathrm{\beta}}),
\end{equation}
where $\Omega(\bm{\mathrm{\beta}})$ is the  the generalized fussed lasso term that applies to $\bm{\mathrm{\beta}}$. Mathematically, $\Omega(\bm{\mathrm{\beta}})$ is defined as
\begin{equation}\label{gfl}
    \Omega(\bm{\mathrm{\beta}}):= \lambda_1 \sum_{(i,j)\in E}|\beta_{ij}| + \lambda_2 \sum_{\substack{(i,j)\in E,\\ (j,k)\in E}} |\beta_{ij}-\beta_{jk}|.
\end{equation}
 The $\sum_{i,j}|\beta_{ij}|$ term aims to remove less important edges in $G$ by enforcing $\beta_{ij}$ to be zero. The $\sum_{(i,j),(j,k)\in E} |\beta_{ij}-\beta_{jk}|$ term promotes smoothness between $\beta_{ij}$ and $\beta_{jk}$ if they have a share vertex $j$. This term makes sure that after removing edges from $G$, the remaining edges can still form connected subgraphs in $G$. The effect of the generalized fussed lasso term is illustrated in Fig.~\ref{JT}(c). 

 For a vertex $i$, if all $\beta_{i,j}=0, \forall (i,j)\in E$, we will remove its vertex embedding in~\eqref{molfeature} by setting the corresponding $w_i=0$.

\subsection{Prioritizing shared structures in positive samples}
In positive samples (drugs that are known to bind to the target protein), if certain connections between the same chemical substructures frequently appear, it suggests that such connections are important for the positive samples. We would enforce such connections to be preserved rather than removed by our GNN model. Therefore, we add additional terms to our learning objective.
\begin{equation}\label{obj3}
    \min_{\bm{\mathrm{\beta}}, \Theta}: \mathcal{L}(\bm{\mathrm{\beta}}, \Theta) + \eta \|\Theta\|^2 + \Omega(\bm{\mathrm{\beta}})+\Gamma(\bm{\mathrm{\beta}}),
\end{equation}
where $\Gamma(\bm{\mathrm{\beta}})$ is defined as
\begin{equation}\label{Gamma}
\Gamma(\bm{\mathrm{\beta}}):=-\lambda_3\sum_{(i,j)\in E} m_{ij}\beta_{ij}^2.
\end{equation}
$m_{ij}$ is the frequency of the connections between chemical substructure $i$ and chemical substructure $j$ appeared in the positive samples. We note that the larger $m_{ij}$ is, the more penalty is applied to $\beta_{ij}$ to make it away from zero. Therefore, adding $\Gamma(\bm{\mathrm{\beta}})$ in~\eqref{obj3} would preserve connections between chemical substructures that frequently appeared in the positive samples. 

\subsection{Optimization}
The final learning objective of our GNN model is written in~\eqref{obj3}. We first rearrange the terms in~\eqref{obj3} and rewrite it as
\begin{equation}\label{objfinal}
    \min_{\bm{\mathrm{\beta}}, \Theta}: \mathcal{F}(\bm{\mathrm{\beta}}, \Theta) + \eta \|\Theta\|^2 + \Omega(\bm{\mathrm{\beta}}),
\end{equation}
where $\mathcal{F}(\bm{\mathrm{\beta}}, \Theta) = \mathcal{L}(\bm{\mathrm{\beta}}, \Theta)+\Gamma(\bm{\mathrm{\beta}})$ is a smooth function for both $\bm{\mathrm{\beta}}$ and $\Theta$. The reason we write our final learning objective as~\eqref{objfinal} is that it is in the format for the proximal alternative linearized minimization (PALM) algorithm~\cite{Bolte2014}.  The PALM algorithm
applies the proximal forward–backward algorithm~\cite{Bolte2014} to optimize both $\bm{\mathrm{\beta}}$ and $\Theta$ in an alternative manner. Specifically, at iteration $k$, the proximal forward-backward mappings of $\eta \|\Theta\|^2$ and $\Omega(\bm{\mathrm{\beta}})$ for given $\Theta^k$ and $\bm{\mathrm{\beta}}^k$ are  
\begin{subequations}
\begin{align}
&\Theta^{k+1}  \in \min_{\Theta}: \left \{ \frac{a^{k}}{2}\left \|\Theta - U^{k}\right \|_F^2+ \eta \|\Theta\|^2 \right \}; \label{v1}\\
&\bm{\mathrm{\beta}}^{k+1}  \in \min_{\bm{\mathrm{\beta}}}: \left \{ \frac{b^{k}}{2}\left \|\bm{\mathrm{\beta}} - V^{k}\right \|_F^2+ \Omega(\bm{\mathrm{\beta}})\right \} \label{v2}, 
\end{align}
\end{subequations}
where $U^k=\Theta^k-\frac{1}{a^k}\nabla_{\Theta}\mathcal{F} (\Theta^k,  \bm{\mathrm{\beta}}^k)$ and $V^k=\bm{\mathrm{\beta}}^k-\frac{1}{b^k}\nabla_{\bm{\mathrm{\beta}}}F(\Theta^{k+1},  \bm{\mathrm{\beta}}^k )$. $a^k$ and $b^k$ are positive real numbers and $\nabla_{\Theta}\mathcal{F}(\Theta^k,  \bm{\mathrm{\beta}}^k )$ is the derivative of $\mathcal{F}(\Theta,  \bm{\mathrm{\beta}}^k )$ with respect to $\Theta$ at point $\Theta^k$ for fixed $\bm{\mathrm{\beta}}^k$ and $\nabla_{\bm{\mathrm{\beta}}}\mathcal{F}(\Theta^{k+1},  \bm{\mathrm{\beta}}^k )$ is the derivative of $\mathcal{F} (\Theta^{k+1},  \bm{\mathrm{\beta}} )$ with respect to $\bm{\mathrm{\beta}}$ at point $\bm{\mathrm{\beta}}^k$ for fixed $\Theta^{k+1}$. 

For the first proximal forward–backward mapping~\eqref{v1}, we can easily find the closed-form solution (leave the derivation to the audience). For the second proximal forward–backward mapping~\eqref{v2}, we use the algorithm proposed in~\cite{10.1145/2847421} to find its optimal solution. We iteratively solve~\eqref{v1} and~\eqref{v2} to obtain a sequence $\{(\Theta^k, \bm{\mathrm{\beta}}^k)\}_{k\in \mathbb{N}}$. It has been proven that the sequence $\{(\Theta^k, \bm{\mathrm{\beta}}^k)\}_{k\in \mathbb{N}}$ generated by PALM converges to a critical point~\cite{Bolte2014}.

\subsection{Related works}

We briefly discuss some current efforts for constructing explainable GNN models. Despite using  different computational strategies, the following methods can be categorized into one kind: the weight ranking-based interpretation method. They \cite{doi:10.1021/acs.jmedchem.9b00959, 10.5555/3495724.3497370, Ying2019-wb} have been tested on BBBP, MUTAG, Tox21, synthetic datasets, etc. 

AttentiveFP~\cite{doi:10.1021/acs.jmedchem.9b00959} stacks Graph Attention (GAT) layer~\cite{velickovic2018graph} on message passing neural network. Its interpretation comes from the attention weight of the GAT layer. GNNExplainer~\cite{Ying2019-wb} and PGExplainer~\cite{Ying2019-wb} explain the model using a learnable binary feature selector. After training, the interpretation can be obtained by analyzing the values in the binary feature selector (values toward 1 means more important). 

\begin{figure*}[h]
\begin{center}
\includegraphics[width=\textwidth]{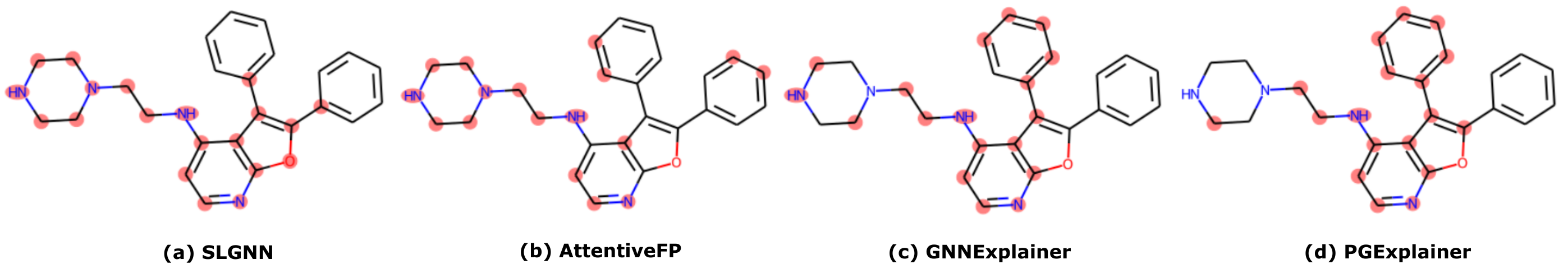}
\end{center}
\caption{The key structures identified by different methods on 2OF2(PDB ID). (a) The atoms in the structure identified by \SLGNN ~are connected and chemically valid. (b) The atoms selected by AttentiveFP are not connected. (c) The atoms selected by GNNExplainer are not connected. (d) The atoms selected by PGExplainer are not connected. } \label{complexs} 
\end{figure*}

The output of the current explainable GNN models is the ranking of the atoms in drug molecules. Therefore, a cutoff is needed to distinguish between important atoms and less important atoms. For tasks like aromatic detection, we can easily select the cutoff~\cite{doi:10.1021/acs.jmedchem.9b00959}. However, for the drug-protein binding task that we focus on in this paper, it is very challenging to pick the cutoff because how many substructures in drug molecules play critical roles in forming the bindings is unknown before chemical experiments.

\section{Experimental Results}
In this section, we evaluate our \SLGNN ~and the competing methods on the drug-protein binding task, which categorize drugs into binding and not binding classes for the target protein. We use \SLGNN ~and the competing methods to build binary classifiers for Lyn, Lck, and Src proteins. We focus on comparing the explanatory power of the competing methods. We first check the predictive power of the key structures identified by the competing methods.  Last but not least, for some known drug-protein bindings, we check whether the chemical substructures identified by the competing methods contain  known binding sites. 



\subsection{Datasets}\label{dataset}
We extract drugs/molecules and their binding information for proteins Lyn, Lck, and Src from the ChEMBL database \cite{Davies2015-qt}. We use the datasets of these three proteins in our experiments. We use Ki values provided in ChEMBL to label the drugs/molecules. 
We consider a drug/molecule with Ki $<$ 1000 nm as a binding class and a drug/molecule with Ki $\geq$ 1000 nm as a not binding class. In sum, for protein Lyn, we extract 649 drugs/molecules, where there are 219 positive samples (binders) and 430 negative samples (non-binders). For protein Lck and Src, we extract data with the same method and get 239 drugs/molecules (93 positive samples and 146 negative samples) and 499 drugs/molecules (218 positive samples and 281 negative samples), respectively.


\subsection{Competing methods}
We consider the drug-protein binding prediction task as a binary classification task. We aim to build models to categorize drugs into binding and not binding classes. Therefore, we use all the competing methods to build binary classifiers.

The competing methods we consider include our \SLGNN and three state-of-the-art methods: AttentiveFP~\cite{doi:10.1021/acs.jmedchem.9b00959}, GNNExplainer~\cite{Ying2019-wb} and PGExplainer~\cite{10.5555/3495724.3497370}.  AttentiveFP is a graph attention-based method whose interpretation is derived from attention weights.  GNNExplainer is a model-agnostic interpretation method based on mean field variational approximation. Given a trained GNN model, it will generate a graph mask for the explanation of this instance.  PGExplainer is built based on a generative probabilistic model, which generates model explanations for new instances without learning from scratch. We use all the models to build binary classifiers for the drug-protein binding task.

\subsection{Hyper-parameters }
For AttentiveFP \cite{doi:10.1021/acs.jmedchem.9b00959}, GNNExplainer \cite{Ying2019-wb} and PGExplainer \cite{10.5555/3495724.3497370}, we use the hyper-parameters suggested in the orginal papers. For our \SLGNN, we set $\lambda_1=1$, $\lambda_2=9$, and $\lambda_3=0.01$ for all experiments.

\subsection{Select the key chemical structures}

For identification of the chemical structures in drugs,  our \SLGNN ~can directly output the key chemical substructures in a drug. However, AttentiveFP \cite{doi:10.1021/acs.jmedchem.9b00959}, GNNExplainer \cite{Ying2019-wb} and PGExplainer \cite{10.5555/3495724.3497370} only generate the rank of the atoms in a drug. To make sure the key structures identified by different methods are comparable, for each drug, we count the number of atoms in the output of \SLGNN ~and use the ranks generated in AttentiveFP \cite{doi:10.1021/acs.jmedchem.9b00959}, GNNExplainer \cite{Ying2019-wb} and PGExplainer \cite{10.5555/3495724.3497370} to pick the same number of atoms in the same drug. Therefore, for any drug used in the learning task, the structures identified by all the competing methods have the same number of atoms.




\subsection{Evaluation of whether the identified structures are chemically valid}

In this section, we check whether the structures identified by the competing methods are chemically valid. 

First, we check whether the structures identified by the competing methods are connected. Fig.~\ref{complexs} shows an example of the structures identified by different methods for the same drug 2OF2\cite{Martin2007} in the Lyn experiment. As illustrated, only the structure identified by our \SLGNN ~is connected. Furthermore, to quantify the connectivity of the  structures identified by the competing methods, we compute the average number of connected components (standard deviation in the parenthesis of Table.~\ref{sec35}) in the structures identified by the competing methods in each drug. Table.1 exhibits such statistics. Clearly, our \SLGNN ~achieves the smallest average connected components, indicating the structures identified by our \SLGNN ~are more connected than the competing methods.

\begin{figure*}[htbp]
\begin{center}
\includegraphics[width=\textwidth]{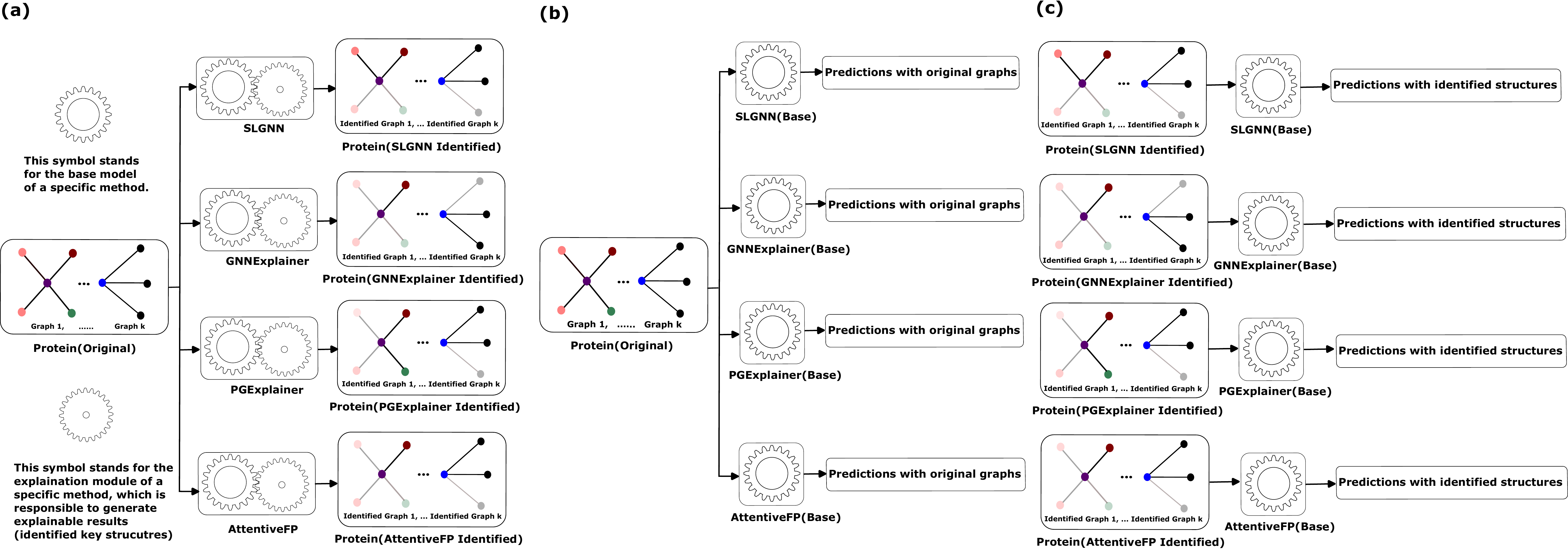}
\end{center}
\caption{In this figure, we illustrate the workflow of experiments in section 3.6. Protein(Original) represents the original Lyn, Lck, Src data. Protein(Method identified) represents the identified key structures for Lyn, Lck and Src data with the corresponding method. In (a) we illustrate how to generate the dataset with identified structures for each competing method. In (b) we introduce how to generate the results for predictions on the original dataset, which correspond to the results on the left panel of Table 2. In (c) we illustrate how to generate the results for predictions on the identified datasets, which correspond to the results on the right panel of Table 2. }
\label{s36}
\end{figure*}

In addition, we checked whether the structures identified by the competing methods are chemically valid. Table.~\ref{sec35} illustrates, for each competing method, the number of drugs whose identified structures are chemically valid. Due to the usage of the chemical-substructure-based graph, all structures identified by our \SLGNN ~are chemically valid. However, for other competing methods, they barely identify chemically valid structures. This experiment demonstrates that the explanatory results generated by the current explainable GNN models are not chemically interpretable. In contrast, our \SLGNN ~overcomes the problem by identifying key structures that consist of chemically valid chemical substructures.  

\begin{table}[h] 
\centering
\begin{adjustbox}{max width=1.0\textwidth}
\begin{tabular}{llll}
\hline Dataset & Model & Avg.CC &  Valid Molecules \\\hline
\multirow{4}{*}{Lyn} & SLGNN & 1.22(0.65) & 649/649 \\
& GNNExplainer & 1.61(0.53) & 0/649 \\
& PGExplainer & 1.76(0.78) & 0/649 \\
& AttentiveFP & 2.42(0.51) & 0/649 \\
\hline
\multirow{4}{*}{Lck} & SLGNN & 1.48(0.78) & 239/239 \\
& GNNExplainer & 1.59(0.67) & 12/239 \\
& PGExplainer & 1.77(0.80) & 13/239 \\
& AttentiveFP & 2.85(1.40) & 14/239 \\
\hline
\multirow{4}{*}{Src} & SLGNN &  1.26(0.76) & 499/499 \\
& GNNExplainer & 1.62(0.42) & 4/499 \\
& PGExplainer & 1.74(0.64) & 8/499 \\
& AttentiveFP & 2.13(0.21) & 6/499 \\
\hline
\end{tabular}
\end{adjustbox}
\caption{Comparison of the average number of connected components, and molecules whose identified structures are all valid. Avg.CC stands for the average number of connected components, and the number after the slash is the total number of molecules in each dataset.} \label{sec35}
\vspace{-0.72cm}
\end{table}

\begin{table*}[htbp]
\centering
\begin{adjustbox}{max width=1.0\textwidth}
\begin{tabular}{lllllllll}
\hline Original Data & & & & Data with identified key structures \\ \hline
\hline Data Used & Model & AUPR & F1 & Data Used & Model & AUPR & F1 \\\hline
Lyn(Original)& SLGNN(Base) & \textbf{0.969(0.045)} & \textbf{0.973(0.027)} & Lyn(SLGNN Identified) & SLGNN(Base) &  \textbf{0.902(0.022)} & \textbf{0.943(0.027)}  \\
Lyn(Original)& GNNExplainer(Base) & 0.843(0.009) & 0.918(0.004) & Lyn(GNNExplainer
Identified) & GNNExplainer(Base) & 0.733(0.021) & 0.786(0.009)\\
Lyn(Original)& PGExplainer(Base) & 0.843(0.009) & 0.918(0.004) & Lyn(PGExplainer
Identified) & PGExplainer(Base)& 0.756(0.018) & 0.799(0.0084)\\
Lyn(Original)& AttentiveFP(Base) & 0.833(0.007) & 0.863(0.008) & Lyn(AttentiveFP
Identified) & AttentiveFP(Base)& 0.695(0.007) & 0.773(0.002)\\
\hline
Lck(Original) & SLGNN(Base) &  \textbf{0.936(0.047)} & \textbf{0.962(0.009)} &  Lck(SLGNN Identified) & SLGNN(Base) & \textbf{0.865(0.024)} & \textbf{0.911(0.033)}\\
Lck(Original) & GNNExplainer(Base) & 0.772(0.041) & 0.836(0.043) &  Lck(GNNExplainer Identified) & GNNExplainer(Base) & 0.700(0.029) & 0.758(0.063)\\
Lck(Original) & PGExplainer(Base) & 0.772(0.041) & 0.836(0.043) &  Lck(PGExplainer Identified) & PGExplainer(Base) & 0.721(0.043) & 0.771(0.041)\\
Lck(Original) & AttentiveFP(Base) & 0.685(0.006) & 0.7472(0.017) &  Lck(AttentiveFP Identified) & AttentiveFP(Base) & 0.632(0.003) & 0.696(0.027)\\
\hline
Src(Original)& SLGNN(Base) &  \textbf{0.974(0.023)} &  \textbf{0.992( 0.012)} &  Src(SLGNN Identified) & SLGNN(Base) & \textbf{0.957(0.026)} &  \textbf{0.921(0.035)}\\
Src(Original)& GNNExplainer(Base) & 0.637(0.016) & 0.693(0.007)&  Src(GNNExplainer Identified) & GNNExplainer(Base) & 0.525(0.019) & 0.563(0.019)\\
Src(Original)& PGExplainer(Base) & 0.637(0.016) & 0.693(0.007) &  Src(PGExplainer Identified) & PGExplainer(Base) & 0.622(0.004) & 0.643(0.032)\\
Src(Original)& AttentiveFP(Base) & 0.794(0.010) & 0.842(0.006)&  Src(AttentiveFP Identified) & AttentiveFP(Base) & 0.694(0.0215) & 0.733(0.006)\\

\hline
\end{tabular}
\end{adjustbox}
\caption{The comparison of the predictive power of the key identified structures. Lyn(Original), Lck(Original), and Src(Original) are the data described in section~\ref{dataset}. Lyn(SLGNN Identified), Lyn(GNNExplainer Identified), Lyn(PGExplainer Identified), and
Lyn(AttentiveFP Identified) are data with the key identified structures identified by each method (described in the 3rd paragraph in section~\ref{perdecitivepower}). SLGNN(Base), GNNExplainer(Base), PGExplainer(Base), and AttentiveFP(Base) are the basic models of each method that switches off the explanatory function (described in the 2nd paragraph in section~\ref{perdecitivepower}).} \label{sec36}
\vspace{-0.70cm}
\end{table*}

\subsection{Evaluation of the predictive power of the identified structures}\label{perdecitivepower}

In this section, we evaluate the predictive power of the key structures identified by the competing methods. We check whether applying the basic message-passing algorithm on the key structures identified by the competing methods could still achieve decent learning accuracy. We use AUPR score~\cite{scikit-learn} and F1 score~\cite{scikit-learn} to evaluate the performance of the drug-protein binding prediction task. To help our audience have a better understanding of the procedure, we draw the workflow of this experiment in Figure.~\ref{s36}.

First, for each competing method, we switch off their function to generate explainable results (identify the key structures) and test their performance on the original data, which is shown in Figure.~\ref{sec36}(b). For our \SLGNN, it becomes applying the basic message passing algorithm on the chemical-substructure-based graphs and we name it SLGNN(Base). For GNNExplainer~\cite{Ying2019-wb} and PGExplainer~\cite{10.5555/3495724.3497370}, they become applying the basic message passing algorithm on the atom-based graphs and we name them GNNExplainer(Base) and PGExplainer(Base), respectively. And for AttentiveFP~\cite{doi:10.1021/acs.jmedchem.9b00959}, it becomes applying the basic message passing algorithm and the graph attention to the atom-based graphs and we name it AttentiveFP(Base). We apply SLGNN(Base), GNNExplainer(Base), PGExplainer(Base), and AttentiveFP(Base) to the original Lyn, Lck, and Src data and show their performance in Table.~\ref{sec36} in the left part. Obviously, our SLGNN(Base) achieves the highest learning accuracy for all three learning tasks due to the use of chemical-substructure-based graphs. Surprisingly, using chemical-substructure-based graphs can significantly improve learning accuracy over using atom-based graphs. 

Then, for each competing method, we switch back on their function to generate explainable results, therefore, they can identify the key structures in a drug. And this procedure is illustrated in Figure.~\ref{sec36}(a). We apply all the competing methods to identify key structures for each drug in the original data. Then, for each protein (Lyn, Lck, and Src), we construct four new data sets  where we replace the original structures of the drugs with the corresponding key structures identified by the competing methods. For example, after applying \SLGNN ~to the Lyn data, we construct a new dataset, in which we replace the drug with the corresponding key structures of the drug identified by \SLGNN ~and we call this new data Lyn(\SLGNN ~Identified). Similarly, we can construct Lyn(GNNExplainer Identified), Lyn(PGExplainer Identified), and Lyn(AttentiveFP Identified), respectively. For Lck and Src proteins, we construct four new datasets as for Lyn, respectively.

\begin{table*}[htbp]
\centering
\begin{adjustbox}{max width=1\textwidth}
\begin{tabular}{llllllllllll}
\hline Lyn & & & & Lck & & & & Src & & & \\ \hline
\hline PDB ID & Model & AUPR & F1 &PDB ID & Model & AUPR & F1&PDB ID & Model & AUPR & F1 \\
\hline 
\multirow{4}{*}{2ZV8} & SLGNN & \textbf{0.621} & \textbf{0.268} & \multirow{4}{*}{2OF2} & SLGNN &  \textbf{0.626} & \textbf{0.667} & \multirow{4}{*}{1KSW} & SLGNN &  0.612 & 0.211 \\ 
  & GNNExplainer &  0.587 & 0.266 &  & GNNExplainer &  0.474 & 0.333 &  & GNNExplainer &  0.55 & 0.5 \\
    & PGExplainer &  0.607 & 0.154 &  & PGExplainer &  0 & 0 &  & PGExplainer &  \textbf{0.766} & \textbf{0.588} \\
  & AttentiveFP & 0.565 & 0.143 &  & AttentiveFP &  0.530 & 0.4 &  & AttentiveFP &  0.633 & 0.153 \\
\hline 
\multirow{4}{*}{2ZV9}  & SLGNN  &  0.482 & \textbf{0.545} & \multirow{4}{*}{2OF4} & SLGNN &  0.237 & 0.222 & \multirow{4}{*}{2SRC} & SLGNN &  0.572 & 0.222 \\ 
  &  GNNExplainer  &  0.466 & 0.250 & & GNNExplainer  & 0 & 0 &  & GNNExplainer &  \textbf{0.649} & 0.352 \\ 
  & PGExplainer  &  \textbf{0.599} & 0.333 &  & PGExplainer  & 0 & 0 &  & PGExplainer &  0.647 & \textbf{0.444} \\ 
  & AttentiveFP &  0 & 0 &  & AttentiveFP  & \textbf{0.265} & \textbf{0.285} &  & AttentiveFP &  0 & 0 \\ 
\hline 
\multirow{4}{*}{2ZVA} & SLGNN  &  \textbf{0.556} & \textbf{0.572} & \multirow{4}{*}{3B2W} & SLGNN & \textbf{0.544} & \textbf{0.286} & \multirow{4}{*}{4K11} & SLGNN &  \textbf{0.624} &  \textbf{0.727} \\ 
  & GNNExplainer &  0.536 & 0.222 &  & GNNExplainer  & 0.530 & 0.273 &  & GNNExplainer &  0.45 & 0.224 \\ 
  & PGExplainer  &  0.474 & 0.200 &  & PGExplainer  & 0.534 & 0.2 &  & PGExplainer &  0.60 & 0.333  \\ 
  & AttentiveFP &  0 & 0 &  & AttentiveFP  & 0 & 0 &  & AttentiveFP &  0.433 & 0.363 \\ 
\hline 
\multirow{4}{*}{3A4O} & SLGNN &  0.630 & \textbf{0.533} &      &  &  &  &      &  &  &  \\
& GNNExplainer &  \textbf{0.726} & 0.200 &      &  &  &  &      &  &  &  \\
 & PGExplainer &  \textbf{0.726} & 0.200 &      &  &  &  &      &  &  &  \\
 & AttentiveFP &  0.671 & 0.182 &      &  &  &  &      &  &  &  \\ \hline
\end{tabular}
\end{adjustbox}
\caption{Results for case studies. The comparison on finding the chemical substructures that contain binding sites. We have 10 drug-protein binding tests.  }\label{binding}
\vspace{-0.72cm}
\end{table*}

Last, to test whether the identified key structures have predictive power, for each competing method, we switch off their function to generate explainable results again and apply them to the identified key structures only. This procedure is shown in Figure.~\ref{sec36}(c). Specifically, for Lyn, we apply SLGNN(Base), GNNExplainer(Base), PGExplainer(Base), and AttentiveFP(Base) to Lyn(\SLGNN ~Identified), Lyn(GNNExplainer Identified), Lyn(PGExplainer Identified), and Lyn(AttentiveFP Identified), respectively. We test whether applying the basic message passing algorithm on the key structures identified by the competing methods could still achieve decent learning accuracy. Table.~\ref{sec36} the right part illustrates the results. As shown, applying the message-passing algorithm to the key structures identified by \SLGNN ~could still achieve high learning accuracy (above 0.85 for AUPR and 0.9 for F1). Compared to applying the message-passing algorithm to the original data (left part of Table.~\ref{sec36}), there is not much accuracy loss. However, when applying the message passing algorithm to the key structures identified by other competing methods, the learning accuracy has become much lower (less than 0.8 for both AUPR and F1).  Therefore, the results imply that the key structures identified by \SLGNN ~have more predictive power than the key structures identified by other competing methods.


\vspace{-0.11cm}

\subsection{Case studies}

In this section, we use the known drug-protein binding data as the ground truth to evaluate the explanatory power of each method. The ground truth data we have included is the binding data: For Lyn, the binding between ANP and Lyn (PDB ID 2ZV8~\cite{Williams2009-mr}), PP2 and Lyn (PDB ID 2ZV9~\cite{Williams2009-mr}), 1N1 and Lyn (PDB ID 2ZVA~\cite{Williams2009-mr}), STU and Lyn (PDB ID 3A4O~\cite{Miyano2009-sr}). For Src, NBS and Src (PDB ID 1KSW~\cite{Witucki2002}), ANP and Src (PDB ID 2SRC~\cite{Xu1999}), 0J9 and Src (PDB ID 4K11~\cite{Madej2013}). For Lck, the binding between 547 and Lck (PDB ID 2OF2~\cite{Martin2007}), 979 and Lck (PDB ID 2OF4~\cite{Martin2007}), 9NH and Lck (PDB ID 3B2W~\cite{Deak2008}).  The binding sites between the drugs and the proteins mentioned above are calculated and collected from PLIP~\cite{Adasme2021}.

We check whether the chemical substructures identified by each method contain the known binding sites. Given a drug (for example ANP) and a protein (for example Lyn), we use the known binding data to classify all the chemical substructures in the drug into two classes: chemical substructures that contain at least a binding site are assigned in the positive class and chemical substructures that do not contain a binding site are assigned in the negative class. The labels of these two classes serve as our ground truth. We consider each method as a binary classifier, which identifies key structures in the drug. For each method, the chemical substructures in the identified key structures are categorized in the positive class and the rest are categorized in the negative class. We then evaluate each method by comparing its classification results against the ground truth. For each drug-protein binding pair, we will use AUPR score~\cite{scikit-learn} and F1 score~\cite{scikit-learn} to evaluate each competing method. 

We want to mention that the key structures identified by AttentiveFP, GNNExplainer, and PGExplainer are not always valid chemical substructures. Therefore, we remove the atoms in the key structures identified by these methods if they do not form a valid chemical substructure. However, for our \SLGNN, we guarantee that the identified key structures consist of valid chemical substructures.

Table.~\ref{binding} illustrates the comparison between the competing methods. As shown in the left part of Table.~\ref{binding}, our \SLGNN ~outperforms other methods in terms of AUPR and F1 for finding the chemical substructures that contain binding sites for protein Lyn except for the AUPR for 3A4O. The outperformance of our \SLGNN ~denotes that most of the chemical substructures identified by \SLGNN ~bind to the protein Lyn. As shown in the middle part of Table.~\ref{binding}, we have the comparison on finding the chemical substructures that contain binding sites for protein Lck. For 2OF2, \SLGNN ~outperforms other methods. For 2OF4, \SLGNN ~and AttentiveFP have competitive performance. AttentiveFP obtains slightly better AUPR and F1 scores. For 3B2W, GNNExplainer attains the best AUPR score but \SLGNN ~achieves the best F1 score. As shown in the right part of Table.~\ref{binding}, we have the comparison on finding the chemical substructures that contain binding sites for protein Src. We notice that for drug 4K11, our \SLGNN ~outperforms other methods. However, for 1KSW and 2SRC, PGExplainer and GNNExplainer achieve better scores.

\begin{table}[htbp]
\centering
\begin{adjustbox}{max width=1.0\textwidth}
\begin{tabular}{lll}
\hline  Model & AUPR &  F1 \\\hline
 SLGNN & \textbf{0.550}(0.114) & \textbf{0.425}(0.192) \\
 GNNExplainer & 0.497(0.184) & 0.262(0.121) \\
 PGExplainer & 0.495(0.260) & 0.245(0.175) \\
 AttentiveFP & 0.310(0.274) & 0.153(0.147) \\
\hline
\end{tabular}
\end{adjustbox}
\caption{Comparison of the average and standard deviation of AUPR and F1 scores of the 10 test cases shown in Table~\ref{binding}.} \label{sec37small}
\vspace{-0.72cm}
\end{table}

In sum, we remark that our \SLGNN ~is the most robust method among the competing methods. We have 10 drug-protein binding tests in Table.~\ref{binding}. \SLGNN ~achieves the best accuracy in 7 out of 10. Even for the ones \SLGNN ~does have the best performance, its performance is close to the best ones. However, for the competing methods, they might perform well on one dataset but their overall performance is not good. As shown in Table.~\ref{sec37small}, we compute the average and standard deviation of the AUPR and F1 scores obtained on the 10 drug-protein binding tests (listed in Table.~\ref{binding}) for each method. Clearly, \SLGNN ~achieves the best average AUPR and F1 scores.

\section{Conclusion}
In this paper, we propose an explainable GNN model called SLGNN. \SLGNN ~is a GNN model built on the chemical-substructure-based graphs, where nodes represent chemically valid substructures. Using chemical-substructure-based graphs is superior to using the widely used atom-based graphs. Because any subgraph in a chemical-substructure-based graph guarantees to be chemically valid. However, the atom-based graphs do not have such nice properties. In addition,  \SLGNN ~applies generalized fussed lasso to the message-passing algorithm to identify key structures of each drug used in the GNN model. The generalized fussed lasso makes sure to the identified structures yielded by \SLGNN ~are connected. We test \SLGNN ~and the state-of-the-art competing methods on three real-world drug-protein binding datasets. We have demonstrated that the key structures identified by our \SLGNN ~are chemically valid and have more predictive power. 

Our \SLGNN ~overcomes the current explainable GNN models' limitations for generating interpretation for the drug-protein binding prediction. These limitations include the explanations generated by explainable GNN models are not always chemically valid and they rely on manual selection of the cutoff threshold that is hard to pick in practice. Our \SLGNN ~resolves the limitations. Therefore,  we believe that our \SLGNN ~has the potential to yield a meaningful explanation for drug-protein binding, which would lead to promising outcomes in drug discovery.

\section*{Acknowledgements}


\bibliographystyle{ACM-Reference-Format}
\bibliography{sample-base}

\end{document}